\documentclass[12pt,preprint]{aastex}
\usepackage{graphicx}
\usepackage{txfonts}
\usepackage[colorlinks=true,citecolor=blue]{hyperref}
\bibpunct{(}{)}{;}{a}{}{,} 
\usepackage[switch,pagewise]{lineno}
\usepackage{subfigure}
\usepackage{overpic}
\usepackage{multirow}
\usepackage{ulem}
\usepackage{color}

\shorttitle{Quasi-periodic wiggles in radio zebra patterns}
\shortauthors{Yu et al.}
\begin{document}

\title{Quasi-periodic wiggles of microwave zebra structures in a solar flare}
\author{Sijie Yu}
\affil{ Key Laboratory of Solar Activity, National Astronomical Observatories Chinese Academy of Sciences, Beijing 100012, China.}
\email{sjyu@nao.cas.cn}
\author{V.~M. Nakariakov}
\affil{Centre for Fusion, Space and Astrophysics, Physics Department, University of Warwick, Coventry, CV4 7AL, UK}
\affil{School of Space Research, Kyung Hee University, Yongin, 446-701, Gyeonggi, Korea}
\affil{Central Astronomical Observatory at Pulkovo of RAS, 196140 St Petersburg, Russia}
\author{L. A. Selzer} 
\affil{Centre for Fusion, Space and Astrophysics, Physics Department, University of Warwick, Coventry, CV4 7AL, UK.}
\author{Baolin Tan and Yihua Yan}
\affil{ Key Laboratory of Solar Activity, National Astronomical Observatories Chinese Academy of Sciences, Beijing 100012, China.}

\begin{abstract}
Quasi-periodic wiggles of microwave zebra pattern structures with period range from about 0.5 s to 1.5 s are found in a X-class solar flare on 2006 December 13 at the 2.6-3.8 GHz with the Chinese Solar Broadband Radio Spectrometer (SBRS/Huairou).  Periodogram and correlation analysis show that the wiggles have two-three significant periodicities and almost in phase between stripes at different frequency. The Alfv\'en speed estimated from the zebra pattern structures is about $700\; \mathrm{Km \,s^{-1}}$. We obtain the spatial size of the waveguiding plasma structure to be about 1 Mm with the detected period of about 1 s. It suggests the ZP wiggles can be associated with the fast magnetoacoustic oscillations in the flaring active region. The lack of  a significant phase shift between wiggles of different stripes suggests that the ZP wiggles are caused by a standing sausage oscillation.
\end{abstract}


\section{Introduction}\label{sec-1}
Quasi-periodic pulsations (QPP) are a frequently observed phenomenon in the electromagnetic emission generated by solar and stellar flares in a vast energy range from radio to hard X-ray and gamma-ray bands \citep[see][for recent reviews]{2009SSRv..149..119N,2010PPCF...52l4009N,2010ApJ...723...25T}. Typically, QPP appear as a pronounced oscillatory pattern in the intensity of the radiation, with the typical periods ranging from a fraction of a second to several minutes. 
Also, QPP have been found as oscillations of the Doppler shift of the emission lines associated with the hot plasma in flaring sites \citep{2006ApJ...639..484M} or its density
\citep{2012ApJ...756L..36K}. The variety of the characteristic periods and modulation depths
of QPPs suggests that they can be caused by several different mechanisms, including
wave-particle interaction \citep[e.g.][]{1987SoPh..111..113A}, 
spontaneous or driven periodic regimes of magnetic reconnection  \citep[e.g.][]{1987ApJ...321.1031T,2010PPCF...52l4009N}, magnetohydrodynamic (MHD) oscillations \citep[e.g.][]{2009SSRv..149..119N}, and oscillations in an equivalent LCR circuit \citep[e.g.][]{2008PhyU...51.1123Z}. 
Revealing the mechanisms responsible for the production of QPP remains an important task
in the context of our understanding of the basic physical processes operating in solar and stellar flares.

Another interesting phenomenon of flaring microwave emission is the zebra pattern structures (ZPs) of the broadband spectral observations: sets of almost-parallel stripes superposed on the microwave type II and IV bursts with slowly frequency drifting and variations \citep[e.g.][]{2006SSRv..127..195C}.
A similar phenomenon is an \lq\lq evolving emission line\rq\rq\ (EEL),  that in contrast with ZP consists of one
single emission stripe in the dynamical spectrum \citep{1998A&A...334..314C,2000A&A...364..853N}.   
There is no broadly-accepted interpretation for ZP, although recent observational 
findings favour the model associating ZP with the coherent generation of 
upper hybrid waves at multiple double plasma resonances in a non-uniform plasma
\citep{1975SoPh...44..461Z}. In this model, the frequency separation of the adjacent stripes in a ZP is directly proportional to the electron gyrofrequency and hence to the magnetic field strength.
This property provides one with a unique method for measuring the coronal magnetic field. 
The model based upon the double plasma resonance is supported by some observational
evidence \citep[e.g.][]{2003A&A...410.1011Z,2007SoPh..246..431C, 2011ApJ...736...64C,2012ApJ...761..136Y}. 
However, other proposed mechanisms, e.g. based upon the whistler wave packets \citep[e.g.][]{2006SSRv..127..195C} 
and trapped upper-hybrid Z-mode waves \citep{2003ApJ...593.1195L} have not been ruled out.
In addition, very recently \cite{2013A&A...552A..90K} proposed a new model that links ZP with propagating compressive MHD waves.
 
Analysis of some ZPs indicates the presence of periodic modulation.
In particular,  \cite{2005A&A...437.1047C} found that intensity of ZP stripes observed on 21 April 2002
pulsated quasi-periodically: the bright ZP stripes consisted of separate short-duration pulses with the period of about 30~ms. Pulsations of the intensity in the adjacent stripes were found to be similar. The detected periodicity was associated with the oscillatory nonlinear
interaction of whistlers with ion-sound and Langmuir waves.  
That ZP event obtained great attention, and the quasi-periodic pulsations of the 
intensity have been considered in several follow-up studies that addressed temporal characteristics
of the pulses and their polarisation \citep[e.g.][]{2007SoPh..246..431C,2008SoPh..253..103K}.
\citep{2007SoPh..246..431C} proposed that the pulsations were associated with the relaxation
oscillations in the system of an electron beam and plasma waves. 
\cite{2007SoPh..241..127K} linked the pulsations with the periodic injection of electron beams.
 \cite{2008SoPh..253..103K} interpreted the quasi-periodic patterns in terms of downward propagating fast magnetoacoustic waves. A similar model was recently employed to the interpretation of the phenomenon of fiber bursts in the dynamical spectra of flare-generated radio emission \citep{2013A&A...550A...1K}. Variations of the intensity with longer periods, about 275~ms, were detected in the
 event on 15 April 1998 by \cite{2000A&A...364..853N}.

Another, less-studied type of ZP modulation is the periodic quasi-coherent oscillating drift 
of the spectral stripes, also called \lq\lq wiggling\rq\rq.
In the unusually-long radio event on 17 February 1992 observed with ARTEMIS, OSRA and IZMIRAN in the band 100-500~MHz, \cite{1998A&A...334..314C} detected pronounced wiggling
of ZP with the period of about 3 min, with the frequency variation amplitude of about 5~MHz.
The relative amplitude of the spectral variation was 2\%. It was linked with the possible variation of the
emitting plasma density by about 4\% or of the magnetic field by 2\%, or a combination of both.
An evolving emission line observed simultaneously with the ZP, showed a similar variability.
In the event of 15 April 1998 observed with Huairou at about 3~GHz, \cite{2000A&A...364..853N}, \cite{2000PASJ...52..919N}
 and \cite{2001ChJAA...1..525C} found that central frequencies of the stripes fluctuated on a typical time scale of 0.5~s and 1.5--2 s in two different time intervals. 
In the shorter-period case, three stripes were found to wiggle synchronously by about 200 MHz with the relative frequency variation of about 6\%. It was estimated to correspond to the relative variation of the magnetic field of 10\% or the density of about 6\%.  In the longer-period case, the spectral amplitude of the oscillations was 80~MHz, from 3.41~GHz to 3.49~MHz. A similar wiggling evolution of ZP stripes can be seen in Fig. ~8 of \cite{2011ApJ...736...64C}, that has not been analysed in detail yet. 
Visual inspection of the figures gives that the period is about 0.3~s and the amplitude about 20 MHz, that is about 1.5\% of the central frequency of 1.35~GHz.

Generally, the periods detected in ZP fine structure wiggling coincide, by the order of magnitude, with the transverse fast magnetoacoustic crossing time in a typical loop of a coronal active region 
\citep[e.g.][]{DeMoortelNakariakov}. Moreover, the required
amplitudes of the variations of the magnetic field and/or the plasma density, of a few percent, 
are consistent with those observed in these waves in other bands \citep[e.g.][]{2001MNRAS.326..428W}. Thus, it is natural to expect that the periodicity may be associated with either an impulsively-generated fast magnetoacoustic wave train 
\citep[e.g.][]{1984ApJ...279..857R,2004MNRAS.349..705N}
or with a standing sausage mode of a fast magnetoacoustic resonator \citep[e.g.][]{2007AstL...33..706K, 2008PhyU...51.1123Z, 2012ApJ...761..134N}, 
or result from a passage of a perpendicular fast wave through a randomly-structured coronal plasma \citep{2005SSRv..121..115N}. All these mechanisms are of great interest for MHD coronal seismology,
as they bring us the unique information about fine, un-resolved structuring of the coronal plasma, and
can rarely be studied in the EUV band because of its insufficient time resolution.

The aim of this paper is to perform a detailed study of the 
quasi-periodic wiggles in a microwave zebra pattern observed a solar flare.
In Section~\ref{sec-2} the instrumentation and the data analysed are described.
In Section~\ref{sec-3} we present the findings that are discussed in Section~\ref{sec-4}.
 
\begin{figure}[htbp!]
\epsscale{0.95}
\plotone{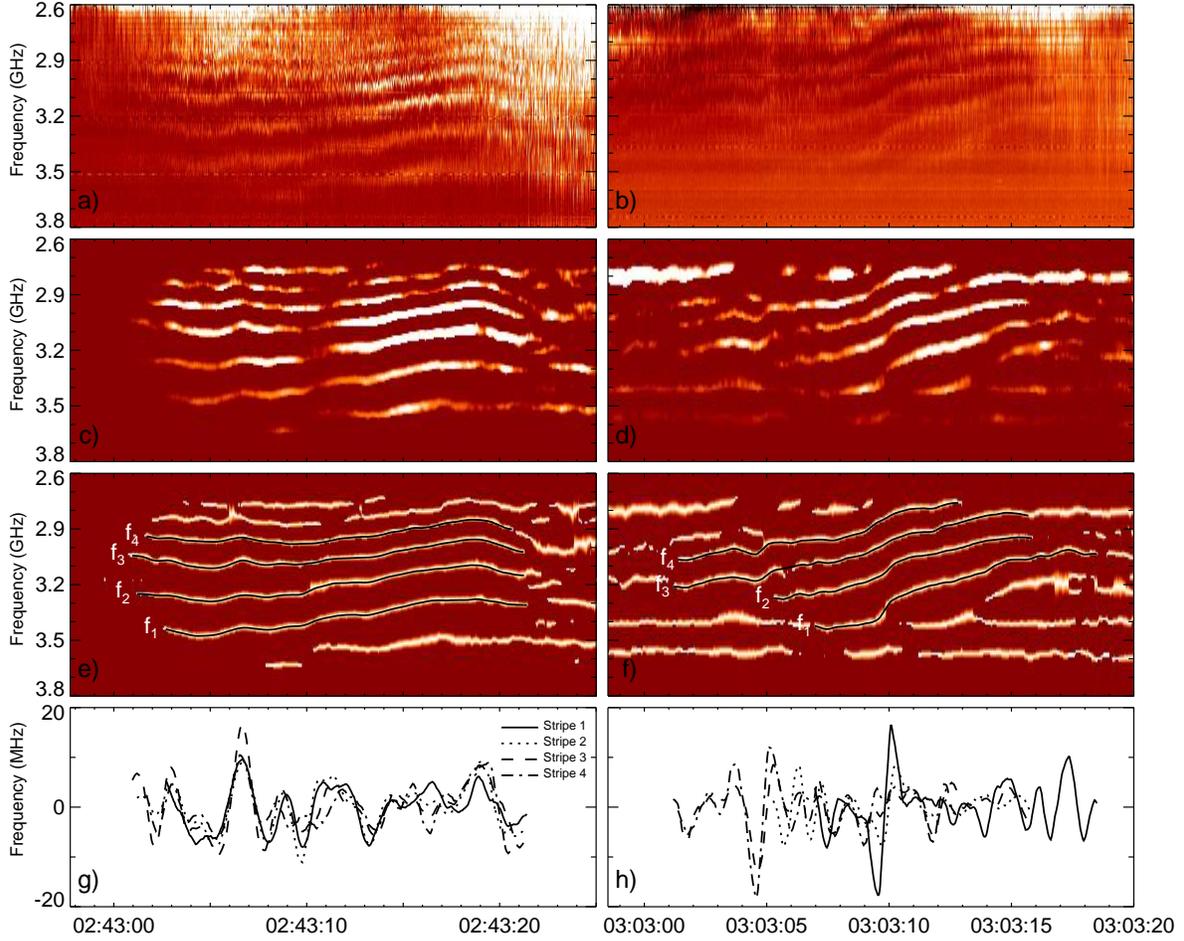}
\caption{Zebra pattern structures on 13 December 2006 observed by  \textit{SBRS/Huairou} at 2.6--3.8 GHz, and the illustration of the processes of extracting the zebra pattern stripes. Panels (a, b) The raw spectrograms at 2.6--3.8GHz on LHCP; (c, d) high contrast images;
(e, f) the rescaled images with the extracted stripes superposed;
(g, h) the detrended stripes frequency $f_\mathrm{N}$. \label{fig-0}}
\end{figure}

\begin{figure}[htbp!]
\epsscale{0.95}
\plotone{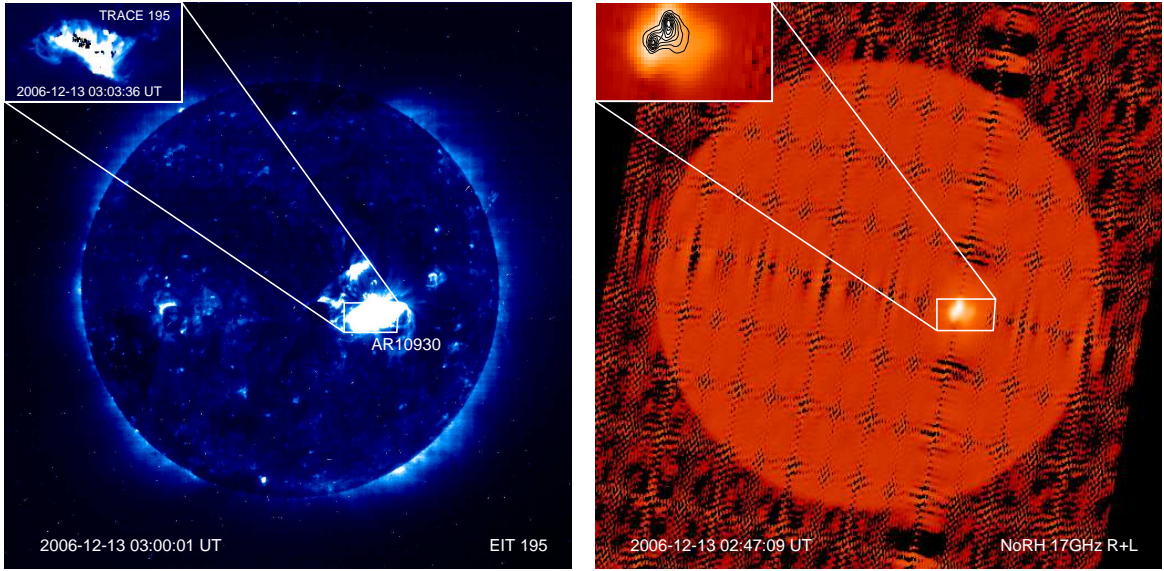}
\caption{ Full disk \textit{SOHO/EIT} 195 \AA\  at 03:00:01 UT and NoRH 17~GHz intensity images at 02:47:09 UT on 13 December 2006. 
Insets show the \textit{TRACE} 195 \AA\ image and the enlarged NoRH 17~GHz image of  AR~10930.  \label{fig-1}}
\end{figure}

\begin{figure}[htbp!]
\epsscale{0.9}
\plotone{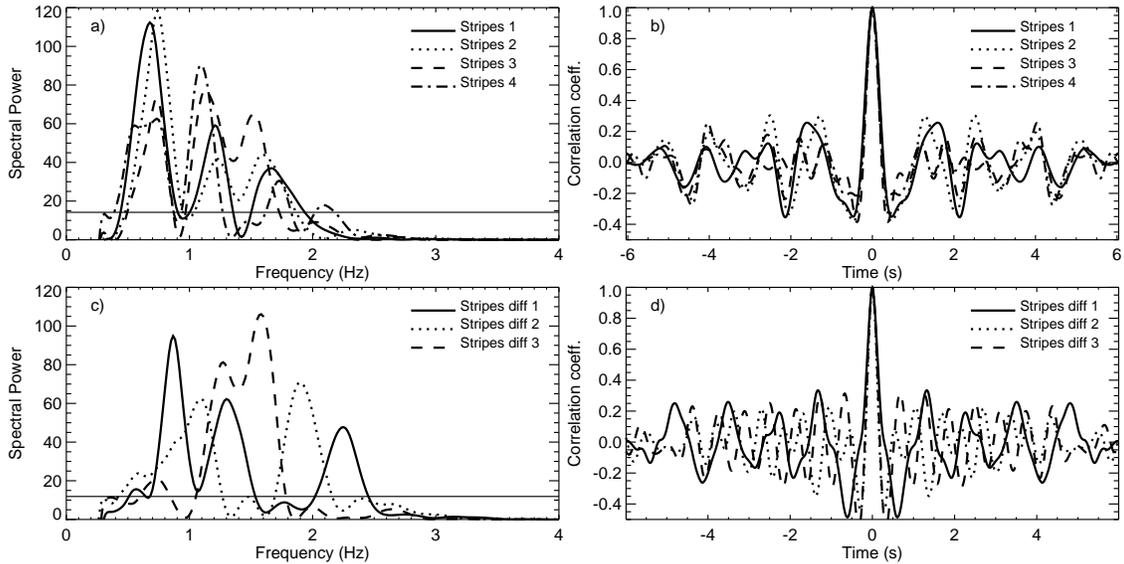}
\caption{Periodograms and auto-correlation functions of QPP components of a,b) the stripe frequency $f_\mathrm{N}$,  and c,d) the frequency separation $\Delta f_\mathrm{N}$ of two neighbouring stripes in the ZP1. The horizontal lines in periodograms indicate the $99.99\%$ confidence level. \label{fig-2}}
\end{figure}

\begin{figure}[htbp!]
\epsscale{0.9}
\plotone{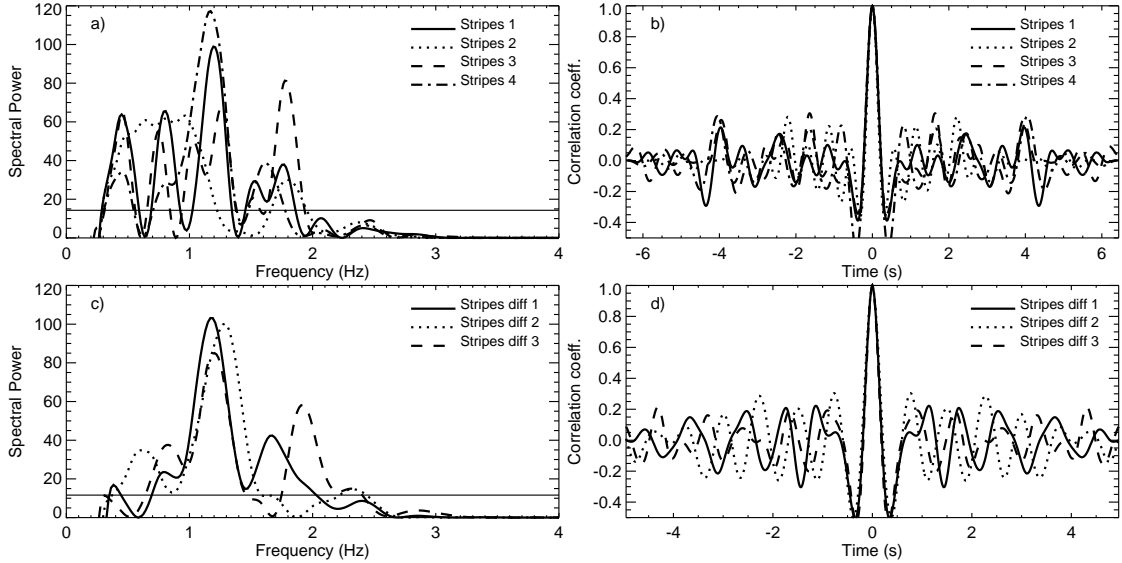}
\caption{The same as in Fig.~\ref{fig-2} for the ZP2. \label{fig-3}}
\end{figure}

\section{OBSERVATIONAL DATA}\label{sec-2}
The flare analysed here occurred on 2006 December 13 in the NOAA active region 10930 located on the disk (S05W33). It was a typical two-ribbon flare that reached the GOES level X3.4/4B class at about 02:40~UT \citep{2007PASJ...59S.807I,2007PASJ...59S.815Y}. This flare was observed by \textit{SOHO}, \textit{TRACE}, \textit{Hinode} and \textit{RHESSI} satellites. 
It was also well observed by the ground-based Chinese Solar Broadband Radio Spectrometer (SBRS/Huairou) \citep{2007ApJ...671..964T, 2007PASJ...59S.815Y} and the Nobeyama Radioheliograph (NoRH). SBRS/Huairou \citep{1995SoPh..160...97F, 2004SoPh..222..167F, 2002ESASP.506..375Y} is a robust solar radio spectrometer that measures the total flux density of solar microwave emission on both left- and right-handed circular polarization (LHCP and RHCP) at three frequency bands: 1.10--2.06 GHz (time resolution of 5 ms and frequency resolution of 4 MHz), 2.6--3.8 GHz (8 ms and 10 MHz), and 5.20--7.60 GHz (5 ms and 20 MHz). NoRH can provide imaging observations at frequency of 17 GHz and 34 GHz \citep{1994IEEEP..82..705N}. Thirteen zebra pattern structures (ZP) were recorded by SBRS/Huairou  at 2.6--3.8 GHz during the flaring process in the discussed event \citep{2012ApJ...761..136Y}. Here we focus on the time interval 02:40--03:05 UT after the soft X-ray emission maximum. Two long-lasting ZPs were detected at 02:43:00--02:43:20 UT (ZP1) and 03:03:00--03:03:20 UT (ZP2) respectively, which show quasi-periodic spectral wiggling of ZP stripes (see Figure~\ref{fig-0}(a)-(b)).

The left panel in Figure \ref{fig-1} shows the full disk EUV image at 195 \AA\ obtained by the Extreme-ultraviolet Imaging Telescope \citep[EIT;][]{1995SoPh..162..291D} onboard \textit{SOHO} during the decay phase of the flare, with the inset of the \textit{TRACE} \citep{1999SoPh..187..229H} image at 195 \AA\  showing the post-flare loop arcade with explosive features at the time of ZP2. The right panel shows the NoRH 17~GHz full disk intensity image, and inset of the enlarged 17~GHz image superposed with a 34~GHz image. The source region of the radio emission at 17~GHz has approximately the same position as the EUV arcade structure in the flare region during the decay phase, with the maximum situated in the northeast of the arcade, between the opposite footpoints of 34~GHz radio sources. The spatial separation of the footpoints is about 50~arcsec. The NoRH 17 GHz full disk image shows that AR 10930 is the unique strong radio emission source on the solar disk in this flare event, indicating that the radiation of the ZPs possibly comes from the flare core region.

The extended duration ($\mathrm{t>15\,s}$) of ZP1 and ZP2 makes it possible to study their long-term variation, including wiggling of the individual spectral stripes. To analyse the oscillatory patterns in ZPs, we need to extract the ZP stripes from the background emission in the raw microwave spectrogram. Specific steps of the data processing are illustrated in Figure \ref{fig-0}. The microwave dynamic spectrograms of ZP1 and ZP2 in the LHCP are shown in Fig.~\ref{fig-0}(a)-(b).  Note that the RHCP spectrograms are not used, for large saturation in the low-frequency range ($<$ 2.9~GHz). The first step of processing is to remove the trend of the background emission to make the bright stripe-like features prominent. For that a running average smoothed over 10 pixels was subtracted from the frequency profile at each instant of time.
We subsequently apply a low-pass filter to smooth out the separate spike-like structures in the stripes.  Then we use the thresholding method to segment the stripe features, shown in Fig.~\ref{fig-0}(c)-(d). The following step is the data series extraction. We fit a Gaussian to the frequency profile of each stripe and then normalised the Gaussians to their amplitudes, so that the brightness of stripes was uniform. This procedure removes the information about the amplitude modulation of the signal, while highlights the frequency modulation. The centres of the Gaussian peaks give us the instant radio-frequency of the stripes. After that, we track the frequency data series of several longest stripes that have been identified manually. 
The variation of the centres of the best-fitted Gaussians in time allows us to obtain the \lq\lq skeletons" 
$f_\mathrm{N}(t)$ of the frequency modulation of the individual spectral stripes in the analysed ZPs ($\mathrm{N}=1, 2, 3 ...$). 
Note that here $N$ denotes the $N$-th extracted stripe enumerated from the highest observed frequency to low frequency in the spectrogram.
Also, one can study the time variation of the difference $\Delta f_\mathrm{N}=f_\mathrm{N+1}-f_\mathrm{N}$ between the neighbouring stripes.
Fig.~\ref{fig-0}(e)-(f) shows the rescaled Gaussian image superimposed with the spectral skeletons of the four stripes of highest frequencies. Note that as the third extracted stripe in ZP2 was seen to be gapped in the raw spectrogram (Fig.~\ref{fig-0}(b)) and the high contrast image (Fig.~\ref{fig-0}(d)) at 03:03:05 -- 03:03:06 UT, and we needed to connect the two segments with a straight line.

\section{RESULTS}\label{sec-3}
The presence of wiggling oscillatory patterns in the two microwave ZPs is 
well seen in the time variation of the spectral skeletons. 
To quantify this finding, we performed the periodogram and autocorrelation analyses of time profiles of $f_\mathrm{N}$ 
and $\Delta f_\mathrm{N}$ of the extracted stripes. The time profiles of $f_\mathrm{N}$ were first smoothed by 30 points ($0.24\,\mathrm{s}$) to remove high-frequency noises,  and then detrended by subtracting the signal $f_\mathrm{N}$ smoothed with a 100 points ($0.8\,\mathrm{s}$) boxcar. The detrended time profiles were shown in Fig.~\ref{fig-0}(g)-(h). Note the smoothing of $f_\mathrm{N}$ attenuates the signal amplitude of oscillation. The oscillation amplitude of the $f_\mathrm{N}$ without smoothing are about 20 MHz, larger than the frequency resolution of the spectrum (10 MHz).
Power spectra of the pre-processed signals were obtained with the use of the Lomb--Scargle periodogram
\citep[e.g.][]{1982ApJ....263..835R} (see the left panels of Fig.~\ref{fig-2}--\ref{fig-3}). The spectra contain significant peaks, above the $99.99\%$ confidence level. 
Additional confirmation of the significance of the detected oscillations was obtained by the application
of Fisher's randomisation test \citep{1985AJ.....90.2317L,2011A&A...533A.116Y}.
The calculation of the spectral peaks for 200 permutation confirmed that
the significance of the main peaks was greater than $99\%$.
To avoid the appearance of artificial periodicities due to the smoothing procedure, we calculated the periodograms of the time profiles $f_\mathrm{N}$ obtained for a set of noise-removing boxcars (10, 20, 30 points) and trend-removing boxcars (60, 70, 80, 100 points)  \citep[see][for a discussion of this method]{2010SoPh..267..329K}. Positions of pronounced spectral peaks in the periodograms do not show any dependence on the smoothing width, implying that these spectral peaks are not the artifact of the smoothing.

Figure \ref{fig-2}(a) presents the periodograms of the time profiles of $f_\mathrm{N}$ of four highest-frequency stripes in ZP1 with the long-term trend removed. There are two well-pronounced  spectral peaks in the vicinities of $0.70$ and $1.20\,\mathrm{Hz}$ ($P_1\sim1.43\,\mathrm{s}$ and $P_2\sim0.83\,\mathrm{s}$) which are seen in all four stripes. The four auto-correlation functions verse time lag (Figure \ref{fig-2}(b)) show evident  in-phase periodic behaviour over several periods. In the auto-correlation functions, the period $P_1$ is well seen in all stripes, while the period $P_2$ is pronounced only for $f_3$ and $f_4$. Figure \ref{fig-2}(c) presents the periodograms of time profiles of detrended spectral difference of the stripes $\Delta f_\mathrm{N}$. The three profiles of $\Delta f_\mathrm{N}$ show different periodic behaviours with three dominant peaks in the vicinities of $0.86$, $1.58$ and $1.90\,\mathrm{Hz}$ ($1.16$, $0.63$ and $0.53\,\mathrm{s}$).  In Figure \ref{fig-2}(d), the auto-correlation functions of detrended $\Delta f_\mathrm{s}$  show less-pronounced periodic oscillatory patterns than obtained for the central frequencies of the stripes, while still significant. The spectral differences $\Delta f_\mathrm{3, 4}$ between the second and third stripes and the third and the fourth stripes (dashed and dotted curves, respectively) have the periods that are apparently two times shorter than the difference between the first and the second stripes 
$\Delta f_\mathrm{2}$.

Figure \ref{fig-3}(a) shows the periodograms of the detrended time profiles of $f_\mathrm{N}$ of the four highest-frequency stripes in ZP2. Two pronounced common peaks are seen in the vicinities of $1.20$ and $1.70\,\mathrm{Hz}$ ($P_2\sim0.83\,\mathrm{s}$ and $P_3\sim0.59\,\mathrm{s}$). The periodicity of $P_2$ is not detected in the detrended time profile of $f_2$. The auto-correlation functions of the detrended time profiles of $f_\mathrm{N}$ all have pronounced oscillatory patterns over several periods(Figure \ref{fig-3}(b)). Periods $P_2$ and $P_3$ are present in the auto-correlation function of $f_1$, $f_3$ and $f_4$. Figure \ref{fig-3}(c) presents the periodograms of time profiles of detrended $\Delta f_\mathrm{N}$, showing a obvious peak at $1.20\,\mathrm{Hz}$ ($P_2\sim0.83\,\mathrm{s}$). The auto-correlation functions in Figure \ref{fig-3}(d) have a similar periodicity for all the three $\Delta f_\mathrm{N}$. 

Additionally, as we can see in Figure \ref{fig-2}--\ref{fig-3}, the periodograms and the auto-correlation functions of the detrended time profiles of $f_\mathrm{N}$ and $\Delta f_\mathrm{N}$ in ZP1 and ZP2 all have a well-pronounced common spectral peak at the period $P_2\sim0.83\,\mathrm{s}$.
The observed amplitude of the frequency variation in the ZP wiggles is about
20 MHz, that is about 0.7\% of the central frequency. 

To establish phase relations between the periodic wiggles of neighbouring stripes, we calculate the cross-correlation coefficients of the detrended time profiles of $f_\mathrm{N}$.
For ZP1,  the highest cross-correlation coefficients, 0.86 between stripes 1 and 2, 0.85 between stripes 2 and 3 and 0.85 between stripes 3 and 4, are obtained for the time lags less than 0.01~s. 
Likewise, for ZP2, the highest cross-correlation coefficients are 0.79 between stripes 1 and 2, 0.56 between stripes 2 and 3 and 0.84 between stripes 3 and 4, for the time lags 0.12~s, -0.02~s and -0.02~s, respectively.
Thus, we can accept that the ZP wiggles occur almost in phase, as in all cases the time lag is found to be much smaller than the oscillation period and are likely to be attributed to noise. 

\section{DISCUSSIONS AND CONCLUSION}\label{sec-4}

Analysis of the fine spectral structure of individual stripes in two microwave ZP observed with 
SBRS/Huairou showed that central frequencies of the stripes performed quasi-periodic oscillations (wiggling) in the range from about 0.5~s to 1.5~s. Simultaneously, the oscillations are found to have two-three significant periodicities. Similar periodicities are detected in the spectral difference of neighbouring ZP stripes. The frequency variation amplitude is about 20~MHz, giving the relative amplitude of 0.7\%.
Both the wiggling periods and amplitudes are consistent with the previous reports of this effect
in \citep{1998A&A...334..314C,2000A&A...364..853N,2000PASJ...52..919N,2001ChJAA...1..525C}.
Wiggling of neighbouring stripes are found to be almost in phase.

The detected periods are of the order of the transverse Alfv\'en or fast magnetoacoustic transit time in active region loops and other plasma non-uniformities (for typical spatial scale of 1~Mm and Alfv\'en speed of 1~Mm/s). This time scale plays an important role in the physics of MHD wave interaction with a structured plasma.
In particular, an impulsively-excited fast-magnetoacoustic disturbance of a coronal loop or a current sheet develops in a quasi-periodic wave train with the period about the transverse fast magnetoacoustic transit time
\citep[][]{1984ApJ...279..857R, 2004MNRAS.349..705N,  2012A&A...546A..49J}.
Also, this time scale is a typical period of standing sausage fast magnetoacoustic modes of such
a loop in the leaky regime \citep[][]{2007AstL...33..706K, 2012ApJ...761..134N}. Thus, it is reasonable to assume that the observed periodicities could be connected with the MHD wave dynamics.

Consider the double plasma resonance (DPR) model \citep[see, e.g.][]{1975SoPh...44..461Z, 2007SoPh..241..127K} as the mechanism responsible for the generation of ZP. According to the DPR model, the microwave ZP structure is likely to be interpreted as a great enhancement of electrostatic upper-hybrid waves at certain resonance levels where the upper hybrid frequency $f_\mathrm{uh}$ is equal to the harmonics of electron cyclotron frequency $f_\mathrm{ce}$:
\begin{equation}\label{equ-0}
f_\mathrm{uh}=(f_\mathrm{pe}^{2}+f_\mathrm{ce}^{2})^{1/2} \simeq {s}f_\mathrm{ce}
\end{equation}
where $f_\mathrm{pe}$ is the plasma frequency of electrons, and ${s}$ is an integer harmonic number. 
We would like to stress that the index $s$ used here is different from the index $N$ used above. Enumeration of the observed ZP stripes with the index $N$ begins from the stripe of the highest observed frequency, while 
some higher-frequency stripes can be missing. Hence, $N=s-M$, where $M$ is the number of missing stripes.

If the plasma density and the absolute value of the magnetic field, and hence the electron plasma and cyclotron frequencies, vary with height, there are several spatially-separated levels where the DPR condition is satisfied. Radio emission from different DPR levels come at the local upper hybrid frequencies. 
The emission from different DPR levels form different individual stripes of the ZP structure. 
When taking into account that $f_\mathrm{pe}\gg f_\mathrm{ce}$, i.e. the upper hybrid frequency $f_\mathrm{uh} \simeq f_\mathrm{pe}$, the emission frequency $f_\mathrm{s}$ of a ZP stripe at the harmonics $s$ equals to the plasma frequency $f_\mathrm{pe}$ or its harmonic. The frequency separation between the neighbouring ZP stripes at the harmonics $\mathrm{s}$ and $\mathrm{s+1}$ is 
\begin{equation}\label{equ-1}
\Delta f_\mathrm{s}=f_\mathrm{s+1}-f_\mathrm{s} \simeq \frac{m}{1-(2L_\mathrm{n}/L_\mathrm{B})}f_\mathrm{ce},
\end{equation}
where  $L_\mathrm{n} = n_\mathrm{e} (\partial n_\mathrm{e}/\partial h)^{-1}$ and $L_\mathrm{B} = {B} (\partial {B}/\partial h)^{-1}$ at the height $h$; ${B}$ and $n_\mathrm{e}$ are the magnetic field and the plasma density, respectively, that are non-uniform in the vertical direction. Hence we obtain 
\begin{equation}\label{equ-2}
f_\mathrm{s} \simeq mf_\mathrm{pe} \sim n_\mathrm{e}^{1/2},
\end{equation}
\begin{equation}\label{equ-3}
\Delta f_\mathrm{s} \sim mf_\mathrm{ce} \sim \mathrm{B}.
\end{equation}
Here the number $m$ describes the mechanism of wave coalescence. In DPR model, $m=1$ when the emission generates from the coalescence of two excited plasma waves, and the polarization will be very weak; $m=2$ when the emission generates from coalescence of an excited plasma wave and a low frequency electrostatic wave, and the polarization will be strong. Thus,  the time profiles of $f_\mathrm{s}$ and $\Delta f_\mathrm{s}$ of the ZP stripes are associated with the variations of the plasma density and magnetic field.
Both these quantities are perturbed by MHD waves, especially, by standing and propagating sausage modes of coronal plasma structures.

Interpreting the observed ZP in terms of the DPR model, we can estimate the background plasma density and magnetic field. Taking that the frequency of the observed emission, about 3 GHz, is about the electron plasma frequency, we obtain that the concentration of electrons is
$n_{e} = 1.1\times10^{11} \;\mathrm{cm^{-3}}$. In this work, the ZP structure is moderate right polarization, the emission possibly generates from the coalescence of two excited plasma wave ($m=2$), and the plasma density of the corresponding emission frequency ($\sim 3.0 \;\mathrm{GHz}$) is $2.8\times10^{10} \;\mathrm{cm^{-3}}$.
Taking that in the DPR resonant layer the electron cyclotron frequency is s times lower than the emission frequency, one can estimate the value of the magnetic field in the layer. It requires the knowledge of the
harmonic number $s$. 

The harmonic number of a ZP stripe can be calculated if the ZP posses at least three stripes \citep{2012ApJ...761..136Y}. For an $s$-th stripe of and an $(s+i)$-th stripe, the harmonic number is  
\begin{equation}\label{equ-4}
 {s} = \frac{i\, \delta _{\mathrm{s+i}}}{\delta _{\mathrm{s}} - \delta _{\mathrm{s+i}}},
\end{equation}
 where $\delta_\mathrm{s} = \Delta f_\mathrm{s}/f_\mathrm{s}$, $\delta_\mathrm{s+i} = \Delta f_\mathrm{s+i}/f_\mathrm{s+i}$, and $f_\mathrm{s} = f_\mathrm{N}$ is the observed frequency of the stripe.
For the analysed event it has been done by \citet{2012ApJ...761..136Y}. It was obtained that for both ZP1 and ZP2 the harmonic number of the stripe 1 is $s\approx10$. 
This gives us the magnetic field of  $50\;\mathrm{G}$. 

Thus, we estimate the Alfv\'en speed as $C_\mathrm{A} \approx 700 \;\mathrm{km \,s^{-1}}$. 
This value is consistent with the typical estimations of the Alfv\'en speed in solar coronal active regions obtained by the method of MHD coronal seismology \cite[e.g.][]{DeMoortelNakariakov}.
The absence of the spatial resolutions in the observation of ZP in the discussed event, does not allow us to determine the spatial location of the sources of individual ZP stripes.  However, according to the estimations made in\citep{2011ApJ...736...64C} for a similar event when simultaneous imaging and spectroscopic observations over a large bandwidth were available, the sources of neighbouring ZP stripes are separated by about 3~Mm. We may assume that in the discussed event the spatial reparation of the individual sources is of the same order. 
The oscillating wiggles of the ZP stripes can be caused by magnetoacoustic waves that perturb the magnetic field and plasma density. 
Magnatoacoustic waves with short periods are known to be present in the corona, e.g., 
from high spatial and time resolution observations during eclipses \citep{2001MNRAS.326..428W,2003A&A...409..325C}. The waves are present in the corona in standing and propagating forms. 
If the observed ZP wiggling is produced by a propagating wave, 
the detected oscillation period, about 1~s, and the estimated value of the Alfv\'en speed, about $700 \;\mathrm{km \,s^{-1}}$ allow us to estimate the wavelength. 
In particular, phase speeds of fast magnetoacoustic waves guided by coronal plasma non-uniformities are of the order of the Alfv\'en speed in the non-uniformity \citep[e.g.][]{1984ApJ...279..857R}. 
For example, such a quasi-periodic fast wave train could result from an impulsive excitation \citep[e.g.][]{2004MNRAS.349..705N} with the characteristic period given by the ratio of the transverse spatial scale of the waveguiding plasma structure and the Alfv\'en speed.
Thus, the wavelength of propagating fast waves with the period of about 1~s is about 1~Mm. In the case of the slow magnetoacoustic wave, taking into account that in the corona the sound speed is lower than the Alfv\'en speed, the wavelength becomes even shorter than 1~Mm. Thus, if the oscillation was caused by a propagating magnetoacoustic wave, neighbouring ZP stripes, separated by several Mm, would be positioned at different phases of the perturbation. 
Hence, neighbouring ZP stripes would wiggle with a significant phase difference that is not detected. 
Consequently, we rule out the interpretation of the ZP wiggling in terms of propagating waves. 
On the other hand, in a global standing mode all segments of the waveguiding plasma non-uniformity oscillate in phase. Moreover, in a global sausage mode of sufficiently thin field-aligned plasma non-uniformities, the period is determined by the transverse size of the plasma non-uniformity and is almost independent of its longitudinal size
\citep{2007AstL...33..706K,2012ApJ...761..134N}.
Taking that the period of the detected oscillations is 1~s,
we obtain that the required spatial size of the waveguiding plasma non-uniformity with the estimated value of the Alfv\'en speed is about 1~Mm. This value is a typical minor radius of an active region loop.
Consequently, the observed ZP wiggling can be associated with fast magnetoacoustic oscillations in the flaring
active region. Moreover, the established lack of a significant phase shift between oscillations of 
different stripes, that are coming from different spatial locations, indicates that the MHD oscillation is likely to be standing. 
All above suggests that the detected ZP wiggles are caused by a standing sausage oscillation. This conclusion is supported by the finding that both instant frequencies of individual stripes and their spectral separation oscillate with the same periods. It is consistent with a sausage oscillation that perturbs both the plasma density and magnetic field \citep{2012ApJ...761..134N}. More information could be obtained from the analysis of 
phase relationship between instant frequencies of individual stripes and their spectral separation. 
But, such a study requires an observational example of a ZP with higher amplitude wiggling.

\acknowledgments
The work is supported by NSFC Grant No. 11221063, 11273030, MOST Grant No. 2011CB811401, and the National Major Scientific Equipment R\&D Project ZDYZ2009-3. This research was also supported by the Marie Curie PIRSES-GA-2011-295272 \textit{RadioSun} project, the European Research Council under the \textit{SeismoSun} Research Project No. 321141 (VMN), the grant 8524 of the Ministry of Education and Science of the Russian Federation (VMN), and the Kyung Hee University International Scholarship (VMN).


\begin{thebibliography}{}

\bibitem[Aschwanden(1987)]{1987SoPh..111..113A} 
Aschwanden, M.~J.\ 1987, \solphys, 111, 113

\bibitem[Chen et al.(2011)]{2011ApJ...736...64C} 
Chen, B., Bastian, T.~S., Gary, D.~E., \& Jing, J.\ 2011, \apj, 736, 64

\bibitem[Chen \& Yan(2007)]{2007SoPh..246..431C} 
Chen, B., \& Yan, Y.\ 2007, \solphys, 246, 431 

\bibitem[Chernov(2006)]{2006SSRv..127..195C} 
Chernov, G.~P.\ 2006, \ssr, 127, 195 

\bibitem[Chernov et al.(1998)]{1998A&A...334..314C} 
Chernov, G.~P., Markeev, A.~K., Poquerusse, M., et al.\ 1998, \aap, 334, 314 

\bibitem[Chernov et  al.(2005)]{2005A&A...437.1047C} 
Chernov, G.~P., Yan, Y.~H., Fu, Q.~J., \& Tan, C.~M.\ 2005, \aap, 437, 1047 

\bibitem[Chernov et al.(2001)]{2001ChJAA...1..525C} 
Chernov, G.~P., Yasnov, L.~V., Yan, Y.-H., \& Fu, Q.-J.\ 2001, \cjaa, 1, 525

\bibitem[Cooper et al.(2003)]{2003A&A...409..325C} 
Cooper, F.~C., Nakariakov, V.~M., \& Williams, D.~R.\ 2003, \aap, 409, 325 

\bibitem[Delaboudini{\`e}re et al.(1995)]{1995SoPh..162..291D} 
Delaboudini{\`e}re, J.-P., Artzner, G.~E., Brunaud, J., et al.\ 1995, \solphys, 162, 291 

\bibitem[De Moortel \& Nakariakov(2012)]{DeMoortelNakariakov} 
De Moortel, I., \& Nakariakov, V.~M.\ 2012, Royal Society of London Philosophical Transactions Series A, 370, 3193

\bibitem[Fu et al.(1995)]{1995SoPh..160...97F} Fu, Q., Qin, Z., Ji, H., 
\& Pei, L.\ 1995, \solphys, 160, 97 

\bibitem[Fu et al.(2004)]{2004SoPh..222..167F} 
Fu, Q.~J., Ji, H.~Q., Qin, Z.~H., et al.\ 2004, \solphys, 222, 167

\bibitem[Handy et al.(1999)]{1999SoPh..187..229H} 
Handy, B.~N., Acton, L.~W., Kankelborg, C.~C., et al.\ 1999, \solphys, 187, 229 

\bibitem[Isobe et al.(2007)]{2007PASJ...59S.807I} 
Isobe, H., Kubo, M., Minoshima, T., et al.\ 2007, \pasj, 59, 807 

\bibitem[Jel{\'{\i}}nek et al.(2012)]{2012A&A...546A..49J} 
Jel{\'{\i}}nek, P., Karlick{\'y}, M., \& Murawski, K.\ 2012, \aap, 546, A49 

\bibitem[Karlick{\'y}(2013)]{2013A&A...552A..90K} 
Karlick{\'y}, M.\ 2013, \aap, 552, A90 

\bibitem[Karlick{\'y} et al.(2013)]{2013A&A...550A...1K} 
Karlick{\'y}, M., M{\'e}sz{\'a}rosov{\'a}, H., \& Jel{\'{\i}}nek, P.\ 2013, \aap, 550, A1 

\bibitem[Kim et al.(2012)]{2012ApJ...756L..36K} 
Kim, S., Nakariakov, V.~M., \& Shibasaki, K.\ 2012, \apjl, 756, L36

\bibitem[Kopylova et al.(2007)]{2007AstL...33..706K} 
Kopylova, Y.~G., Melnikov, A.~V., Stepanov, A.~V., Tsap, Y.~T., \& Goldvarg, T.~B.\ 2007, Astronomy Letters, 33, 706 

\bibitem[Kupriyanova et al.(2010)]{2010SoPh..267..329K} 
Kupriyanova, E.~G., Melnikov, V.~F., Nakariakov, V.~M., \& Shibasaki, K.\ 2010, \solphys, 267, 329 

\bibitem[Kuznetsov(2008)]{2008SoPh..253..103K} 
Kuznetsov, A.~A.\ 2008, \solphys, 253, 103

\bibitem[Kuznetsov \& Tsap(2007)]{2007SoPh..241..127K} 
Kuznetsov, A.~A., \& Tsap, Y.~T.\ 2007, \solphys, 241, 127 

\bibitem[LaBelle et al.(2003)]{2003ApJ...593.1195L} 
LaBelle, J., Treumann, R.~A., Yoon, P.~H., \& Karlicky, M.\ 2003, \apj, 593, 1195 

\bibitem[Linnell Nemec \& Nemec(1985)]{1985AJ.....90.2317L} 
Linnell Nemec, A.~F., \& Nemec, J.~M.\ 1985, \aj, 90, 2317 

\bibitem[Mariska(2006)]{2006ApJ...639..484M} 
Mariska, J.~T.\ 2006, \apj, 639, 484 

\bibitem[Nakajima et al.(1994)]{1994IEEEP..82..705N} 
Nakajima, H., Nishio, M., Enome, S., et al.\ 1994, IEEE Proceedings, 82, 705 

\bibitem[Nakariakov et al.(2004)]{2004MNRAS.349..705N} 
Nakariakov, V.~M.,  Arber, T.~D., Ault, C.~E., et al.\ 2004, \mnras, 349, 705 

\bibitem[Nakariakov et al.(2012)]{2012ApJ...761..134N} 
Nakariakov, V.~M., Hornsey, C., \& Melnikov, V.~F.\ 2012, \apj, 761, 134 

\bibitem[Nakariakov et al.(2010)]{2010PPCF...52l4009N} 
Nakariakov, V.~M., Inglis, A.~R., Zimovets, I.~V., et al.\ 2010, Plasma Physics and Controlled 
Fusion, 52, 124009 

\bibitem[Nakariakov \& Melnikov(2009)]{2009SSRv..149..119N} 
Nakariakov, V.~M., \& Melnikov, V.~F.\ 2009, \ssr, 149, 119 

\bibitem[Nakariakov et al.(2005)]{2005SSRv..121..115N} 
Nakariakov, V.~M., Pascoe, D.~J., \& Arber, T.~D.\ 2005, \ssr, 121, 115 

\bibitem[Ning et al.(2000a)]{2000A&A...364..853N} 
Ning, Z., Fu, Q., \& Lu, Q.\ 2000, \aap, 364, 853 

\bibitem[Ning et al.(2000b)]{2000PASJ...52..919N} 
Ning, Z., Fu, Q., \& Lu, Q.\ 2000, \pasj, 52, 919 

\bibitem[Roberts et al.(1984)]{1984ApJ...279..857R} 
Roberts, B., Edwin, P.~M., \& Benz, A.~O.\ 1984, \apj, 279, 857 

\bibitem[Scargle(1982)]{1982ApJ....263..835R} 
Scargle, J. D.\ 1982, \apj, 263, 835

\bibitem[Tajima et al.(1987)]{1987ApJ...321.1031T} 
Tajima, T., Sakai, J., Nakajima, H., et al.\ 1987, \apj, 321, 1031

\bibitem[Tan et al.(2007)]{2007ApJ...671..964T} Tan, B., Yan, Y., Tan, C., 
\& Liu, Y.\ 2007, \apj, 671, 964 

\bibitem[Tan et al.(2010)]{2010ApJ...723...25T} 
Tan, B., Zhang, Y., Tan,  C., \& Liu, Y.\ 2010, \apj, 723, 25

\bibitem[Williams et al.(2001)]{2001MNRAS.326..428W} 
Williams, D.~R., Phillips, K.~J.~H., Rudawy, P., et al.\ 2001, \mnras, 326, 428 

\bibitem[Yan et al.(2002)]{2002ESASP.506..375Y} 
Yan, Y.~H., Fu, Q.~J., Liu, Y.~Y., \& Chen, Z.~J.\ 2002, in the 10th European
Solar Physics Meeting, Solar Variability: From Core to Outer Frontiers, Vol. 1, ed.
A. Wilson (ESA SP-506; Noordwijk: ESA Publication Division), 375

\bibitem[Yan et al.(2007)]{2007PASJ...59S.815Y} 
Yan, Y.~H., Huang, J., Chen, B., \& Sakurai, T.\ 2007, \pasj, 59, 815

\bibitem[Yu et al.(2012)]{2012ApJ...761..136Y} 
Yu, S., Yan, Y., \& Tan, B.\ 2012, \apj, 761, 136 

\bibitem[Yuan et al.(2011)]{2011A&A...533A.116Y} 
Yuan, D., Nakariakov, V.~M., Chorley, N., \& Foullon, C.\ 2011, \aap, 533, A116 

\bibitem[Zaitsev \& Stepanov(2008)]{2008PhyU...51.1123Z} 
Zaitsev, V.~V., \& Stepanov, A.~V.\ 2008, Physics Uspekhi, 51, 1123 

\bibitem[Zheleznyakov \& Zlotnik(1975)]{1975SoPh...44..461Z} 
Zheleznyakov, V.~V., \& Zlotnik, E.~Y.\ 1975, \solphys, 44, 461 

\bibitem[Zlotnik et al.(2003)]{2003A&A...410.1011Z} 
Zlotnik, E.~Y., Zaitsev, V.~V., Aurass, H., Mann, G., \& Hofmann, A.\ 2003, \aap, 410, 1011 

\end{thebibliography}
\end{document}